\documentclass[a4paper,11pt]{article}
\pdfoutput=1 

\usepackage{jheppub} 

\usepackage[T1]{fontenc} 

\title{\boldmath Fragmentation of projectile nucleus in carbon-carbon collisions at 4.2 GeV/c per nucleon}

\author[a,1]{J. Shinebayar,\note[*]{Corresponding author.}$^*$}
\author[b,1]{R. Togoo,}
\author[b,2]{Ts. Baatar}
\author[b,3]{B. Otgongerel}
\author[b,4]{M. Sovd}
\author[b,5]{N. Khishigbuyan}
\author[b,6]{and M. Urangua}

\affiliation[a]{Mongolian National University of Education, \\Ulaanbaatar, Mongolia}
\affiliation[b]{Mongolian Academy of Sciences, \\Ulaanbaatar, Mongolia}

\emailAdd{shinebayar@msue.edu.mn}
\emailAdd{togoo@mas.ac.mn}
\emailAdd{baatarts@mas.ac.mn}
\emailAdd{otgongerelb@mas.ac.mn}
\emailAdd{sovdm@mas.ac.mn}
\emailAdd{khishigbuyan\_n@mas.ac.mn}
\emailAdd{urangua\_m@mas.ac.mn}

\abstract{The temperature characteristics of carbon spectator fragments formed in carbon collisions with carbon nuclei at a primary momentum of 4.2 GeV/c per nucleon were presented and discussed on corrected experimental data. As well as studied the multiplicities formed by the spectator protons, deuterons, and tritons in the inelastic nucleus-nucleus interactions. We found that the temperature absorbed by the spectator fragments is dependent on their mass. \\

}

\begin{document} 
\maketitle
\flushbottom

\section{Introduction}
\label{sec:intro}

In this paper, we carried out the study of fragments that did not participate in the nuclear interactions from the image data recorded by the two-meter propane bubble chamber at the High energy laboratory of the Joint Institute of nuclear research in Dubna, Russia.

The first research work related to the fragmentation process was done in 1956 by Wolfgang et al. In the study, in addition to more studied re-excitation modes such as spallation and fission, a new mode called fragmentation was determined by calculating light fragments formed in hadron-nucleus interactions ~\cite{Wolfgang1956}. Perfilov, Lozhkin, and Shamov studied nuclear fission and fragmentation processes using photo emulsion experimental data ~\cite{Perfilov1960}. The research work ~\cite{Poskanzer1970}, assumes a hadron-nuclear interaction; that is observed the highest energy fragments with apparent nuclear temperatures to 20 MeV at 90$^{\circ}$ in the interaction of 5.5 GeV/c proton-uranium. 

In 1972, the fragments produced by the fragmentation of 29 GeV $^{14} \mathrm{N}$ ions in Carbon and hydrogen were investigated ~\cite{Heckman1972}. In the nuclear-nuclear interaction, the nuclear temperature is inferred from the momentum distributions of the fragments and is approximately equal to the projectile nuclear binding energy, indicative of a small energy transfer between the target and fragment ~\cite{Greiner1975}. The projectile fragments spread out in a narrow conical shape with a momentum close to or greater than the initial momentum per nucleon. Experimentally, relativistic fragments with $z \geq 2$ are easily distinguished by ionization. On the other hand, a study work was carried out in 1996 that distinguished singly charged fragments (protons, deuterons, and tritons) based on the inverse of their momentum ~\cite{Belaga1996}.

The purpose of this paper is to study the multiplicities of projectile fragments in inclusive $\mathrm{C+C}$ interactions, at 4.2 GeV/c per nucleon, as well as to compare how their temperature characteristics

\section{Experimental procedure}
\label{sec:experi}
In this article, the experimental data used here were obtained based on processing stereo photographs from the 2-m propane bubble chamber constructed at the High Energy Laboratory of the Joint Institute for Nuclear Research (JINR, Dubna, Russia), placed in a magnetic field of strength 1.5 T, and irradiated with a beam of protons accelerated to a momentum of 4.2 GeV/c at the JINR synchrophasotron ~\cite{Abdrakhmanov1978}, ~\cite{Agakishiev1987}, ~\cite{Agakishiev1989}. The difficult problem of distinguishing high-energy positive pions from protons in the processing of stereo photographs has been solved and the experimental data has been revised ~\cite{Bondarenko1998}. Statistics of the experimental data consist of 34947 carbon nucleus-propane collisions and of which 18889 events were selected by criteria of inelastic carbon-carbon collisions from the entire ensemble of carbon interactions with propane ($\mathrm{C}_{3} \mathrm{H}_{8}$) ~\cite{Togoo2016}. The initial momentum per nucleon is $p_{0}$=4.2 GeV/c. In the work ~\cite{Belaga1996} it was observed that the maximum values of $1/p_{0}$ and $1/2p_{0}$ of protons and deuterons are the same in the distribution of inverse momentum ($1/p$) along the track of relativistic particles with single charged and emission angle $\theta \leq 4^{\circ}$ in $\mathrm{C}$-$\mathrm{C}_{3} \mathrm{H}_{8}$ (propane) collisions. 

\section{Results}

\begin{figure}[tbp]
\centering 
\includegraphics[width=.65\textwidth]{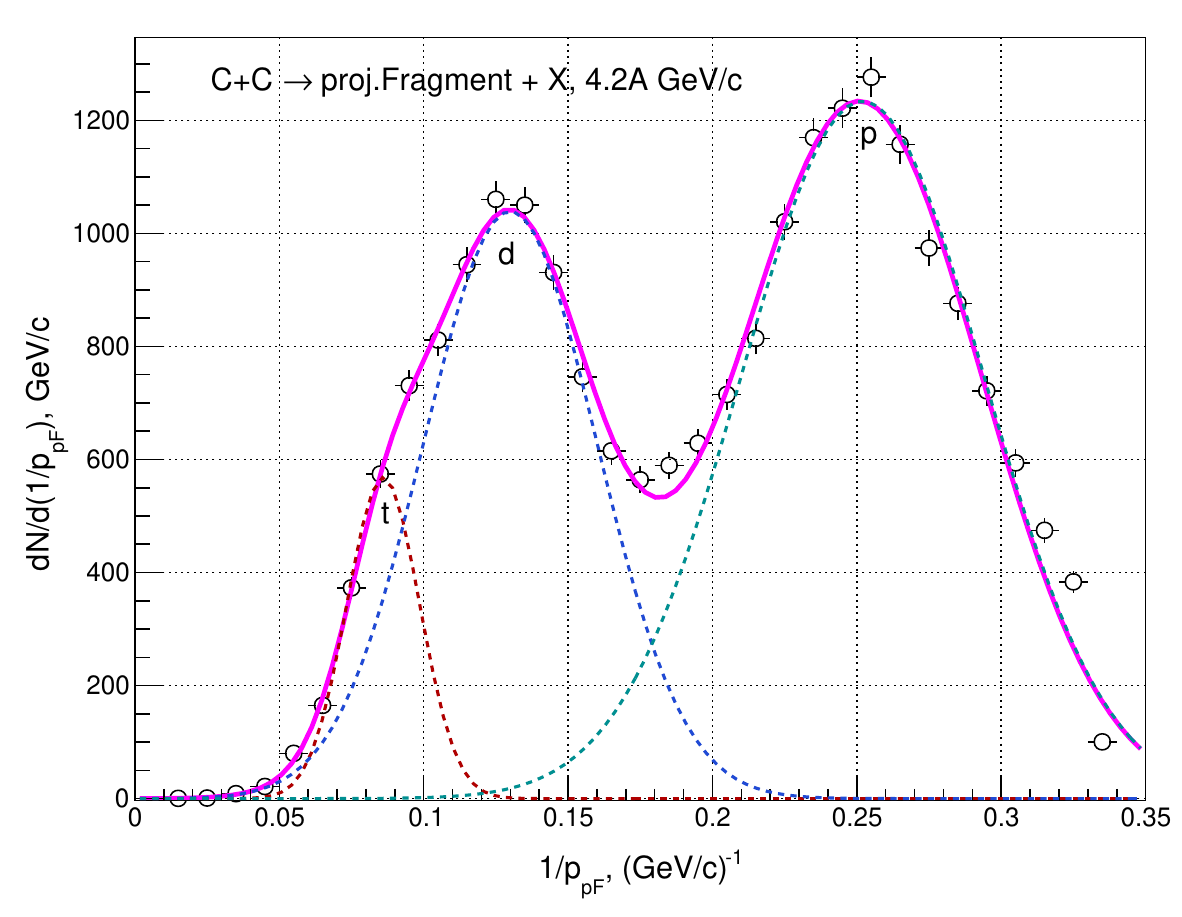}
\caption{\label{fig:1} Distributions of single charged projectile fragment momentum in $\mathrm{C + C}$ collisions. The experimental data is represented by the black circles. The purple curve is a fitting approximation of the sum of three Gaussian functions. The position of the maximum distribution for protons is marked by p (green dashed curve), deuteron by d (blue dashed curve), and triton by t (red dashed curve).}
\end{figure}

In this research work, we have established the inverse momentum distribution of projectile-nucleus with single charged fragments (protons, deuterons, and tritons), (see figure ~\ref{fig:1}). As can be seen from this figure, the spectators limited the momentum of protons in the interval 3 GeV/c < $p$ < 5.9 GeV/c, deuterons in the interval 5.9 GeV/c < $p$ < 10 GeV/c, and tritons in the interval $p$ > 10 GeV/c. If it is defined by the variable 1/$p_{0}$, then 1/$p_{0}$ > 0.17 for protons, 0.1 < 1/$p_{0}$ <0.17 for deuterons, and 1/$p_{0}$ < 0.1 for tritons. The number of fragments outside these momentum intervals must be equal to the number of other single-charged particles within the interval.

In the above intervals, proton, deuteron, and triton multiplicities can be defined. Their average values $\left\langle n_{\mathrm{p}} \right\rangle$, $\left\langle n_{\mathrm{d}} \right\rangle$, and $\left\langle n_{\mathrm{t}} \right\rangle$ are presented in Table ~\ref{tab:1}. On the other hand, the accuracy of momentum measurements in the bubble chamber at $p$ > 10 GeV/c was insufficient for equally confident separation of deuterons from tritons and proton fragments from interacting singly charged particles. The estimated fractions of ``non-fragments'' and deuterons among protons, protons and tritons among deuterons, and, finally, deuterons among tritons selected according to the specified criteria do not exceed 10-25 \% for the groups and subgroups of collisions considered below.

\begin{table}[tbp]
\centering
\begin{tabular}{ccccc}
\hline
 & $\mathrm{p}$ & $\mathrm{d}$ & $\mathrm{t}$ & $\text{PF}(z>1)$\\
\hline 
Experimental data & 0.752 $\pm$ 3 & 0.347 $\pm$ 0.003 & 0.110 $\pm$ 0.004 & 0.746  $\pm$ 0.004\\
\hline
\end{tabular}
\caption{\label{tab:1} Average multiplicities of spectator protons, deuterons, tritons, and multi-charged ($\text{PF}(z > 1)$) fragments in C + C interactions.}
\end{table}

\begin{figure}[tbp]
\centering 
\includegraphics[width=.55\textwidth]{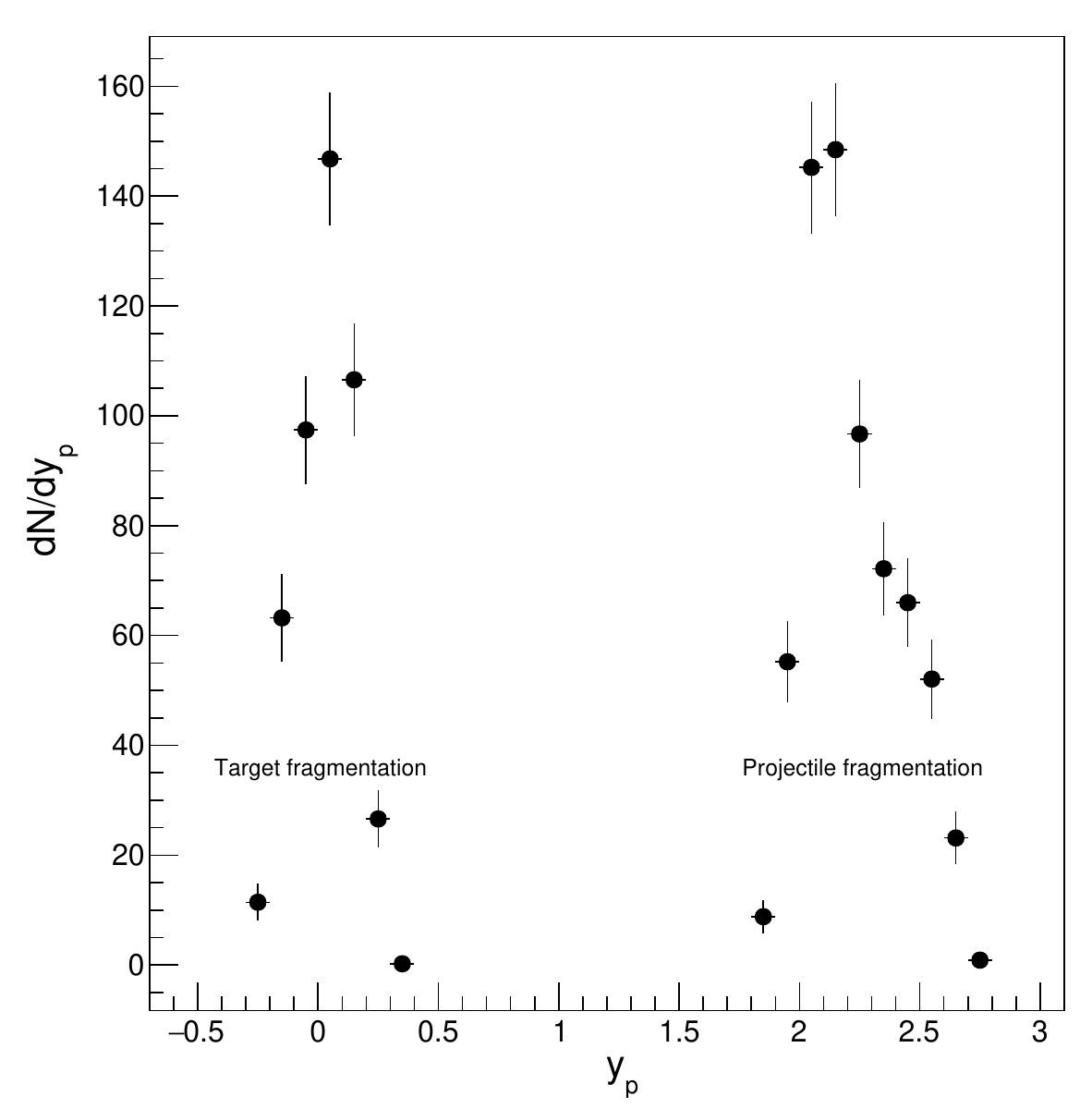}
\caption{\label{fig:2} The longitudinal rapidity distributions for the fragments of a projectile and target nucleus.}
\end{figure}

It can be observed from Figure ~\ref{fig:2}, the protons remaining after the elimination of singly charged fragments of deuterons and tritons of the projectile-carbon nucleus is close to the protons with momenta 150.0 MeV/c > $p$ < 1 GeV/c (target spectators) that are not involved in the inelastic interactions of the target-carbon.

The number of projectiles singly and multiply charged fragments can determine every interaction. For example, we traditionally use $N_{\text{s03}}$ for multiply charged fragments and $N_{\text{sREL}}$ for singly charged fragments, and $N_{\text{PF}}$ (number of projectile fragments) is defined as $N_{\text{PF}}= N_{\text{s03}} + N_{\text{sREL}}$. Here, the first letters of stripping are s, which is REL-relative.

\begin{figure}[tbp]
\centering 
\includegraphics[width=.55\textwidth]{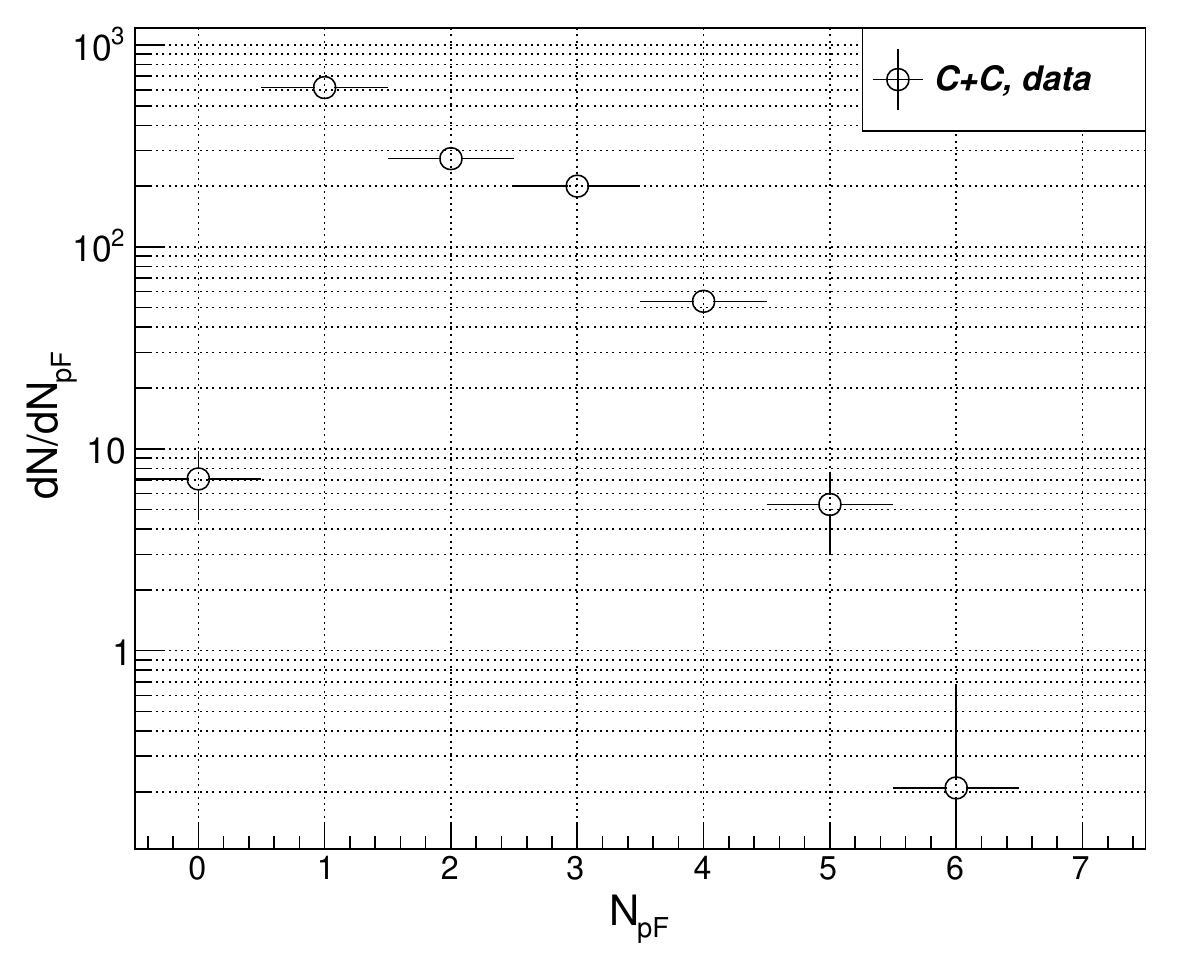}
\caption{\label{fig:3} The distributions of $N_{\text{PF}}$ of projectile-nucleus fragments in experimental data}
\end{figure}

Let's examine how the structure function is distributed based on the transverse mass variable for fragments that are heavier than protons.

\begin{figure}[tbp]
\centering 
\includegraphics[width=.7\textwidth]{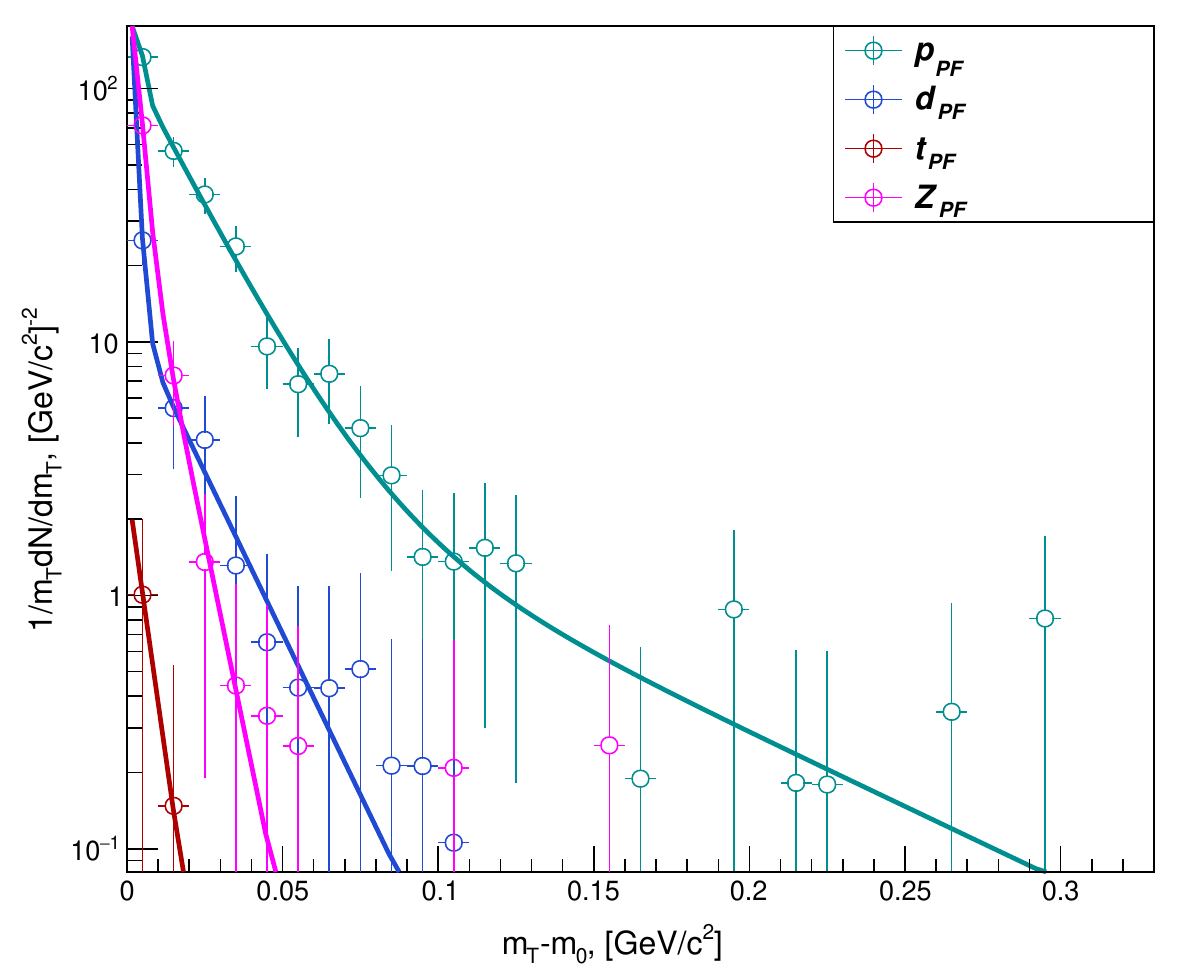}
\caption{\label{fig:4} The variable $m_{\text{T}} (m_{\text{T}}=\sqrt{p_{\text{T}}^{2}+m^{2}})$ distributions as a structure function. Points with statistical errors are experimental data (green-$p_{\text{PF}}$, blue-$d_{\text{PF}}$, red-$t_{\text{PF}}$, purple-$z_{\text{PF}}$). The lines are the fitting by the sum of two or three exponential functions. Where, $m_0$ is the rest mass of a particle.}
\end{figure}

Figure ~\ref{fig:4} illustrates how the sum of three exponential functions can provide an approximation, with the inverse value of their slope parameter representing a quantity similar to temperature. It is represented by the following expression \eqref{eq:1}:
\begin{equation}
\label{eq:1}
\frac{1}{m_{\text{T}}} \frac{\Delta N}{\Delta m_{\text{T}}} \sim a_{1} e^{-m_{\text{T}}/T_{0}^{\text{I}}} + a_{2} e^{-m_{\text{T}}/T_{0}^{\text{II}}} +a_{3} e^{-m_{\text{T}}/T_{0}^{\text{III}}}
\end{equation}

Here, $m_{\text{T}}$ is the transverse mass of the particle. The values of $T_{0}^{\text{I}}$, $T_{0}^{\text{II}}$, and $T_{0}^{\text{III}}$ parameters indicate the presence of two or more phases within the projectile nucleus and are close to the results of ~\cite{Karnaukhov1999}.

Now we consider how the temperature, defined by the $m_{\text{T}}$ variable, depends on the number of projectile fragments (from $N_{\text{PF}}$).

\begin{figure}[tbp]
\centering 
\includegraphics[width=0.9\textwidth]{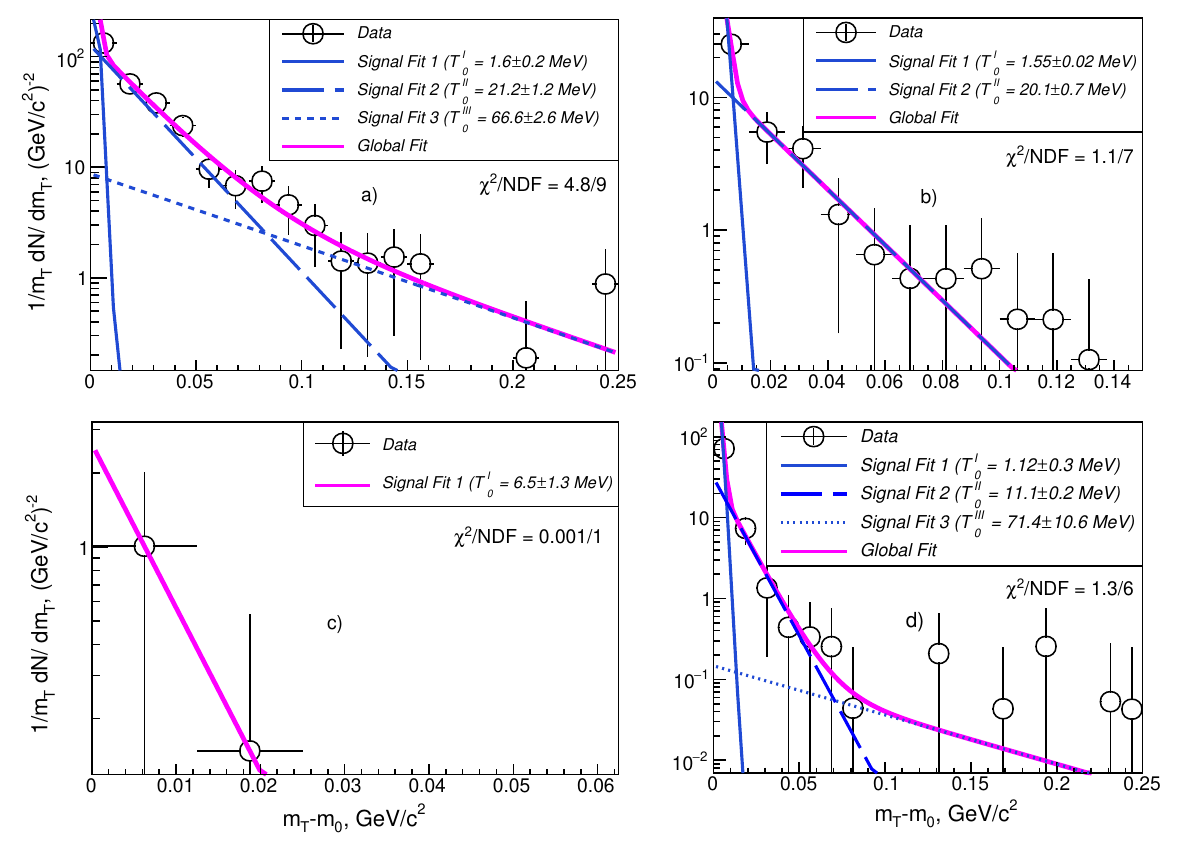}
\caption{\label{fig:5} The variable $m_{\text{T}} (m_{\text{T}}=\sqrt{p_{T}^{2}+m^{2}})$ distributions as structure functions in four values of $N_{\text{PF}}$ (protons-$p_{\text{PF}}$ (a), deuterons-$d_{\text{PF}}$ (b), tritons-$t_{\text{PF}}$ (c) and multi charged fragments $z_{\text{PF}}$ (d)), respectively. The purple lines are the fitting by the sum of two or three exponential functions. Blue and dashed blue lines represent the exponential functions}
\end{figure}

Based on the approximations shown in Figure ~\ref{fig:5}, it can be inferred that the temperature, denoted $T_{0}^{\text{I}}$, $T_{0}^{\text{II}}$, and $T_{0}^{\text{III}}$ remains relatively normal values of temperature. In other words, the temperature characteristics are weakly dependent on the number of fragments heavier than protons. The temperature values are close to those of the FASA experiment, which measures the output of heavy fragments such as deuterons and tritons from the target nucleus ~\cite{Karnaukhov1999}. Temperature values for $p_{\text{PF}}$, $d_{\text{PF}}$, $t_{\text{PF}}$ and $z_{\text{PF}}$ are displayed in Table ~\ref{tab:2}.

\begin{table}[tbp]
\centering
\begin{tabular}{ccccc}
\hline
 & $T_{0}^{\text{I}}$ & $T_{0}^{\text{II}}$ & $T_{0}^{\text{III}}$ & $\chi^{2}/\text{NDF}$\\
\hline 
$p_{\text{PF}}$ & 1.6 $\pm$ 0.2 & 21.2 $\pm$ 1.2 & 66.6 $\pm$ 2.6 & 4.8/9\\
$d_{\text{PF}}$ & 1.55 $\pm$ 0.02 & 20.1 $\pm$ 0.7 & $-$  & 1.1/2\\
$t_{\text{PF}}$ & 6.5 $\pm$ 1.3 & $-$ & $-$ & 0.001/1\\
$z_{\text{PF}}$ & 1.12 $\pm$ 0.3 & 11.1 $\pm$ 0.2 & 71.4 $\pm$ 0.7 & 1.3/6\\
\hline
\end{tabular}
\caption{\label{tab:2} Parameters of the approximation of the exponential function, the values of $T_{0}^{\text{I}}$, $T_{0}^{\text{II}}$, and $T_{0}^{\text{III}}$ in the experiment}
\end{table}

\section{Conclusions}

In conclusion, we have analyzed the projectile spectators in C+C interactions at 4.2 GeV/c per nucleon. The basic conclusions from the research are the following:
\begin{itemize}
\item Single-charged nuclear fragments are classified into protons, deuterons, and tritons using the 1/$p$ variable, and their mass multiplicities were determined. 

\item The temperature values $T_{0}^{\text{I}}$, $T_{0}^{\text{II}}$, $T_{0}^{\text{III}}$ for fragments heavier than protons are estimated by the transverse mass using the structure-function. Our observations confirm the idea of other experimental works that, while considering the nuclear collisions induced by massive beams, the projectile fragments must be ascribed to two emission sources at the least. For multiply charged ($z>1$) projectile fragments, a liquid-gas phase transition determined by the FASA experiment was observed. The temperature characteristics are weakly dependent on the number of fragments produced in each collision.
\end{itemize}

\acknowledgments

We are grateful to the functionaries at Laboratory of High Energies, JINR, Dubna, Russia, for their contribution to the processing of stereo-photographs from a 2-meter propane bubble chamber. We also thank to professor G. Sharkhuu and Ts. Enkhbat Ph. D. for useful consideration, helpful discussions, and valuable advice. This investigation was supported in part by fundamental scientific research of the Mongolian Foundation for Science and Technology (project number, 2019/20 ShuSs).


\end{document}